\documentstyle[aps,multicol,pre,epsfig]{revtex}

\newcommand{\be}{\begin{equation}}
\newcommand{\ee}{\end{equation}}
\newcommand{\lb}{\ell_B}

\newcommand{\sat}{\hbox{\scriptsize sat}}

\newcommand{\Zeff}{Z_{\hbox{\scriptsize eff}}}
\newcommand{\Zbare}{Z_{\hbox{\scriptsize bare}}}

\newcommand{\laeff}{\lambda_{\hbox{\scriptsize eff}}}

\newcommand{\labare}{\lambda_{\hbox{\scriptsize bare}}}
\newcommand{\tz}{t_{\!_Z}}

\begin{document}
\title{Effective charge versus bare charge: an analytical estimate for
colloids in the infinite dilution limit}
\author{Miguel Aubouy$^{1}$, Emmanuel Trizac$^{2}$ and Lyd\'eric Bocquet$^{3}$}
\address{
$^{1}$ SI3M., DRFMC, DSM/CEA-Grenoble, UMR CNRS-CEA-UJF 5819,
17 rue des Martyrs, 38054 Grenoble Cedex 9, France\\
$^{2}$ Laboratoire de Physique Th\'eorique, 
UMR CNRS 8627, B{\^a}timent 210, Universit{\'e}  Paris-Sud,
91405 Orsay Cedex, France\\
$^{3}$ DPM, Universit\'e Claude Bernard Lyon 1, 43 bld du 11 novembre 1918,
69622 Villeurbanne Cedex, France 
}

\date{\today}
\maketitle

\begin{abstract}
We propose an analytical approximation for the dependence of the effective
charge on the bare charge for spherical and cylindrical macro-ions as a
function of the size of the colloid and salt content, for the situation 
of a unique colloid immersed in a sea of electrolyte (where the
definition of an effective charge is non ambiguous).
 Our approach is based
on the Poisson-Boltzmann (PB) mean-field theory. Mathematically speaking,
our estimate is asymptotically exact in the limit $\kappa a\gg 1$,
where $a$ is the radius of the colloid and $\kappa$ the inverse screening
length. In
practice, a careful comparison with effective charges parameters obtained by
numerically solving the full non-linear PB theory proves that
it is good down to $\kappa a\sim 1$. 
This is precisely the limit appropriate to treat colloidal
suspensions. A particular emphasis is put on the range of parameters
suitable to describe both single and double strand DNA molecules under
physiological conditions.
\end{abstract}

\section{Introduction}

We know from the work of Debye and H\"{u}ckel that elementary charges, $e$,
immersed in an electrolyte solution interact through a screened
Coulomb pair potential: $V(r)\sim e^{2}\exp (-\kappa r)/r$, where the
screening length, $\kappa ^{-1}$, characterizes the thermodynamics of the
ionic species. For highly charged macro-ions (bare charge 
$\Zbare e$, where $\Zbare\gg
1 $), the strong electrostatic coupling between the macro-ion and the
micro-ions results in an additional screening so that the actual Coulomb pair
potential still writes $V(r)\sim \Zeff^{2}e^{2}\exp (-\kappa ^{\prime
}r)/r $ at large distances (with possibly $\kappa ^{\prime }\neq \kappa $ 
\cite{Kjellander}), but now $\Zeff$ is an effective (also called
\textquotedblright apparent\textquotedblright , or \textquotedblright
renormalized\textquotedblright ) charge parameter, much smaller in absolute
value than $\Zbare$. The idea is that the charged colloid retains captive a
fraction of the oppositely charged micro-ions in its immediate vicinity, and
therefore apparently behaves as a new entity of lower electric charge.

Because the effective charge is the relevant parameter to compute the
electrostatics of the system at large inter-particle distances, 
the concept of charge renormalization plays a central role in the
thermodynamics of highly charged colloidal suspensions.
Several reviews have appeared recently which discuss this notion \cite%
{Kjellander,Belloni,Hansen,Levin}, see also \cite{Lukatsky}. 
In the colloid science field this concept
has been introduced by Alexander \textit{et al.} \cite{Alexander} in the
context of the Poisson-Boltzmann (PB) cell model, but had been widely
accepted since the fifties in the field of linear polyelectrolytes \cite%
{Katchalsky,Manning}. For an isolated macro-ion on the other hand,
the definition of an effective charge from the far
field potential created in an electrolyte is
unambiguous \cite{Belloni,Letter,JCP,Rque}.

In general the effective charge depends on the geometry of the particle, the
concentration of macro-ions and the thermodynamics of the electrolyte.
Within the non-linear Poisson-Boltzmann (PB) mean-field theory: 
$\Zeff\simeq \Zbare$
for low values of $\Zbare$, and $\Zeff$ saturates to a constant, $\Zeff^{\sat}$%
, when $\Zbare\rightarrow \infty $. Many studies have focused on finding an
approximation for $\Zeff^{\sat}$. In particular, in the limiting
case of infinite dilution,
we have proposed a
matching procedure for the electrostatic potential such that $\Zeff^{\sat}$
may be estimated at the level of the linearized PB theory only 
(the non linearity of the problem occurring only through an effective
boundary condition, see \cite{Letter,JCP}). This
approach may be generalized to account for concentrated suspensions. In
contrast, much less is known about the functional dependence of $\Zeff$ on 
$\Zbare$. This is because finding the exact analytical dependence $\Zeff(\Zbare)$
involves solving the full non-linear PB system of equations, which is out of
reach in general.

Here, we propose an \emph{analytical approximation} for the dependence of $%
\Zeff$ on $\Zbare$ for spherical and cylindrical macro-ions as a function of
the size of the colloid and the salt content. We restrict ourselves to the
infinite dilution limit where an exact analytical representation of
the electrostatic potential, $\psi $, solution of the non-linear PB
theory has been recently obtained \cite{Shkel}. 
Our estimate is asymptotically exact in the
limiting case when $\kappa a\gg 1$. In practice however, it is an accurate
approximation of the exact solution in the whole colloidal domain: $\kappa
a\gtrsim 1$, where $a$ is the typical size of the colloid.

\section{Effective charge dependence}

We consider the situation of an isolated macro-ion of given bare charge in a
symmetrical, mono-valent electrolyte of bulk density $n_{0}$ (no
confinement). The solvent is considered as a medium of uniform dielectric
(CGS) permittivity $\varepsilon $. Within Poisson-Boltzmann theory, 
micro-ions/micro-ions correlations are discarded  and the potential
of mean force identified with the electrostatic potential $\psi$.
Accordingly, the reduced electrostatic potential ($\phi =e\psi /kT$), 
assumed to vanish far from macro-ion, obeys the equation 
\begin{equation}
\nabla ^{2}\phi =\kappa^2 \sinh \phi ,
\end{equation}
where the screening factor $\kappa $ is defined as $\kappa ^{2}=8\pi \ell
_{B}n_{0}$ and the Bjerrum length quantifies the strength of electrostatic
coupling: $\ell _{B}=e^{2}/(\epsilon kT)$ ($kT$ is the thermal energy).
Far away from the colloid, $\phi $ obeys the linearized PB (LPB)
equation $\nabla^2\phi = \kappa^2\phi$. The far field within PB is thus
the same as that found within LPB, provided the charge is suitably 
renormalized is this latter case. 
In particular, for infinite
rods (radius $a$, bare line charge density $\labare e$), we have 
\begin{equation}
\phi (r)\stackrel{r\rightarrow \infty }{\sim }A_{00}\left( \frac{2}{\pi }%
\right) ^{1/2}K_{0}(\kappa r)  \label{phiasymptcyl}
\end{equation}
where $K_{0}$ is the zero order modified Bessel function (the symbol $\sim $
stands for ``asymptotically equal'' ). In
Eq. (\ref{phiasymptcyl}), the prefactor $A_{00}$ is a function of both the
thermodynamics of the electrolyte and the characteristics of the macro-ion: $%
A_{00}(\labare ,\kappa a)$. With these notations, the effective line charge 
density, $\laeff$, is such that 
\begin{equation}
2\ell _{B}\laeff=A_{00}\left( \frac{2}{\pi }\right) ^{1/2}
\kappa a \, \, K_{1}(\kappa a)  \label{effchargeVSAooCyl}
\end{equation}
where $K_{1}$ is the first order modified Bessel function.

Using the method of multiple scales, Shkel \textit{et al} were able to
propose an approximate expression for $A_{00}(\labare ,\kappa a)$ up to the
second order in $(\kappa a)^{-1}\ll 1$ \cite{Shkel}. This result can be
translated into an approximate form for the functional dependence of $%
\laeff(\labare )$ through Eq. (\ref{effchargeVSAooCyl}). After some
algebra, we find, again for cylinders and up to order ${\cal O}(1/(\kappa a))$
\begin{equation}
\laeff \,\lb \, =\,  2\, \kappa a \,\,t_\lambda\, +\, 
\frac{1}{2} \,\left( 5-\frac{t_\lambda^4+3}{t_\lambda^2+1}
\right)\, t_\lambda 
\label{EffchargeCyl}
\end{equation}
where  
\be
t_\lambda = T\left(\frac{\labare \lb}{\kappa a + 1/2}\right)
\ee
and the function $T$, also useful in spherical geometry (see below),
is defined as
\be
T(x) = \frac{\sqrt{1+x^2}-1}{x}.
\label{FonctionT(x)}
\ee
In the limit of diverging bare charge $\labare \to \infty$, 
$t_\lambda \to 1$ so that
the saturation value for the line charge 
density reads 
\begin{equation}
\laeff^{\sat}\ell _{B} =  2\kappa a+ \frac{3}{2} 
+{\cal O}\left(\frac{1}{\kappa a}\right).
\label{EffchargeSatCyl}
\end{equation}
In the opposite limit where $\labare \to 0$, Eq. (\ref{EffchargeCyl})
yields $\laeff = \labare$ as expected. 

Similarly, for spheres (radius $a$, bare charge $\Zbare e$) we find that 
\begin{equation}
\Zeff \,\frac{\lb}{a} \,=\, 4\, \kappa a \,\,\tz\, +\, 
2  \,\left( 5-\frac{\tz^4+3}{\tz^2+1}
\right) \,\tz
\label{EffchargeSph}
\end{equation}
where now 
\be
\tz = T\left(\frac{\Zbare \lb/a}{2\kappa a + 2}\right)
\ee
and the function $T(x)$ is still defined by Eq.
(\ref{FonctionT(x)}).
The corresponding saturation value for the effective charge is 
\begin{equation}
\frac{\ell _{B}}{a}\Zeff^{\sat}=4\kappa a+6  +
{\cal O}\left(\frac{1}{\kappa a}\right),
\label{EffchargeSatSph}
\end{equation}
while for low bare charges, we have
\be
\lim_{\Zbare\to 0} \,\,\,\,\, \frac{\Zeff}{\Zbare} \,\, = \, 1.
\ee

Expressions (\ref{EffchargeCyl})-(\ref{EffchargeSatCyl}) [resp. Eq. (\ref%
{EffchargeSph})-(\ref{EffchargeSatSph})] provide the first analytical
estimate of the functional dependence of the effective charge on the bare
charge for cylindrical (resp. spherical) macro-ions in an electrolyte
solution. They are the exact expansions of the correct result in the limit
of large $\kappa a$. In practice however, they are accurate as soon as $%
\kappa a\gtrsim 1$, as we now show.

\begin{center}
\begin{figure}[h]
\epsfig{figure=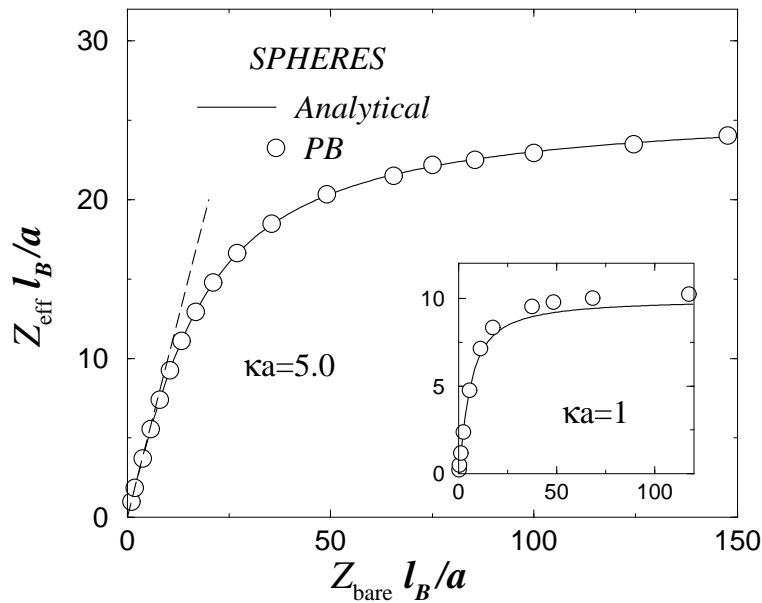,width=10cm,angle=0}
\caption{Rescaled effective charge, $\Zeff\ell_B/a$, \emph{versus}
rescaled bare charge, $\Zbare\ell_B/a$, for isolated spherical macro-ions 
in a symmetrical, monovalent electrolyte solution 
(for $\kappa a=5$, where $\kappa ^{-1}$ is
the Debye-H\"{u}ckel screening length of the electrolyte, and $a$ is the
sphere radius). The solid line is the analytical estimate of $\Zeff$ [see
Eq. (\ref{EffchargeSph})], 
whereas the open circles are the exact values of $\Zeff$
found by numerically solving the full non-linear PB problem. The
dashed line has a of slope unity to emphasize the initial linear
behaviour. The inset shows $\Zeff(\Zbare)$ for $\kappa
a=1$.}
\label{fig:sph}
\end{figure}
\end{center}

\begin{center}
\begin{figure}[h]
\epsfig{figure=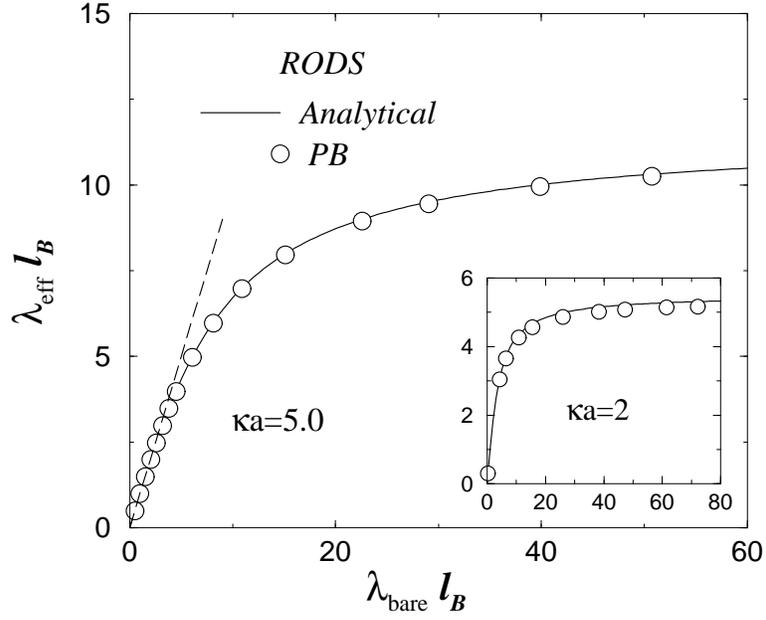,width=10cm,angle=0}
\caption{Effective line charge density, $\laeff$, \emph{versus}  
bare line charge density, $\labare$, for
an isolated, infinitely long cylindrical macro-ion  
($\kappa a=5$). The solid line is the analytical estimate of $\laeff$ [see
Eq. (\ref{EffchargeCyl})], whereas the open circles are the exact values of 
$\laeff$ found by numerically solving the non-linear PB theory 
(the dashed line has a of slope unity). The inset shows 
$\laeff(\labare)$ for $\kappa a=2$.}
\label{fig:cyl}
\end{figure}
\end{center}

We compare on Figs. \ref{fig:sph} and \ref{fig:cyl} 
the results for the effective charge against the
bare charge obtained using: \emph{a)} our estimate and \emph{b)} the exact
solution found by solving the full numerical PB problem. The
results are displayed for two different values of the ionic strength of the
solution. The agreement is seen to be excellent for $\kappa a\gtrsim 5$, and
reasonable down to $\kappa a\approx 1$ (see also below).

\section{Cylindrical geometry and DNA}

\begin{center}
\begin{figure}[h]
\epsfig{figure=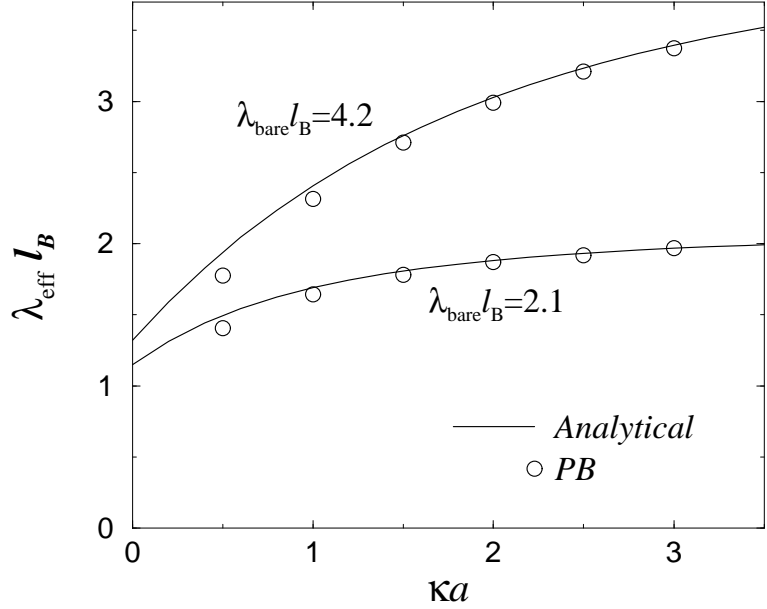,width=10cm,angle=0}
\caption{Effective line charge density, 
$\laeff\ell _{B}$, \emph{versus} $\kappa a$ for isolated, infinitely long
cylindrical macro-ions of bare charge $\labare\ell _{B}=2.1$ (lower sets)
and $\labare\ell _{B}=4.2$ (upper sets). 
The solid line is the analytical estimate of 
$\laeff$ [see Eq. (\ref{EffchargeCyl})], whereas the open circles are the
exact values of $\laeff$ found by numerically solving the full
non-linear PB theory. In the formal limit $\kappa a \to \infty$,
these two curves would saturate to 2.1 and 4.2.}
\label{fig:dna}
\end{figure}
\end{center}

Because DNA (a rod-like polyelectrolyte in first approximation) is of
paramount importance in biology, the cylindrical case deserves a
special attention. For typical parameters of double strand DNA 
($a\cong 10\,$\AA, $\labare \ell _{B}\cong 4.2$), 
the domain $\kappa a\geq 1$, where we
expect our estimate to be good enough, reads $n_{0}\geq 0.1$ $M$ for
monovalent symmetrical electrolyte. This is experimentally relevant. Indeed,
physiological conditions are found for $n_{0}=0.15$ $M$ 
($\kappa ^{-1}=8\,$ \AA). For simple strand DNA 
($a\cong 7\,$\AA , $\labare \ell _{B}\cong 2.1$) 
the condition $\kappa a\geq 1$ writes $%
n_{0}\geq 0.2$ $M$. Fig. \ref{fig:dna} displays the effective charge, 
$\laeff\ell _{B}$, against $\kappa a$ for $\labare \ell _{B}=2.1$, and $%
\labare \ell _{B}=4.2$. We observe that our estimate remains decent
down to $\kappa a = 0.5$, where the error is of the order of 7\%
(see also Tables 1 and 2).

Note that, for DNA in typical experimental conditions, the parameters are
such that we find ourselves in the crossover regime where both the linear
approximation ($\laeff\cong \labare $) and the asymptotic
approximation ($\laeff\cong \laeff^{\sat}$) fail to be
accurate, as shown in Tables 1 and 2. In other words, both $\labare $ and 
$\laeff^{\sat}$ provide a rather poor approximation for $\laeff$. This
justifies \textit{a posteriori} our effort to find a good
approximation for the full functional dependence of 
$\laeff(\labare) $.

\section{Concluding remarks}

In this contribution, we have proposed an analytical approximation for the
dependence of the effective charge on the bare charge for spherical and
cylindrical macro-ions as a function of the size of the colloid and salt
content. Mathematically speaking, our estimate is asymptotically exact in
the limit $\kappa a\gg 1$. In practice, a careful comparison with effective
charges parameters obtained by numerically solving the full non-linear PB
theory proves that it is good down to $\kappa a\gtrsim 1$
(where $a$ is the radius macro-ion). This is precisely the
relevant range of parameters for colloidal suspensions.

An important example considered in some detail is the cylindrical geometry. 
This is
because the infinitely long charged rod provides a simple model for DNA.
While the simple asymptotical approximations fail for DNA in typical
physiological conditions, we provide an estimate for the effective charge
line density, which compares well with the numerical results.

We have performed the analysis at the level of the 
mean-field Poisson-Boltzmann theory.
In spite of its limitations, this picture is excellent 
for all existing macro-ions in water at room temperature,
when only monovalent micro-ions are present in the electrolyte
(see e.g. ref \cite{Levin} for an estimation of a relevant
coupling constant quantifying the importance of the micro-ionic
correlations). It may also hold for multivalent counter-ions,
provided the surface charge density of the macro-ion
is not too large. In general, for a given total charge of the 
macro-ion, micro-ionic correlation
become irrelevant in the asymptotic limit where 
$a\gg\ell_{B}$ \cite{Groot,Stevens}. 




\bigskip\bigskip

\bigskip

\bigskip
\bigskip
\newpage
\noindent \textbf{Table 1: }Comparison of the effective charge and the
saturation value for single strand DNA at physiological
conditions as found \emph{a)} by our analytical estimate, \emph{b)}
numerically solving PB theory.
\medskip

\noindent 
\begin{tabular}{|c|c|c|c|c|c|}
\hline
$\labare \ell _{B}$ & ~~$\kappa a$~~ & 
~~$\laeff\ell _{B}\quad [Eq. (\ref{EffchargeCyl})]$~~ & 
~~$\laeff\ell _{B}\quad (num)$~~ & 
~~$\laeff^{\sat}\ell _{B}\quad [Eq. (\ref{EffchargeSatCyl})]$~~ & 
~~$\laeff^{\sat}\ell _{B}\quad (num)$~~ \\ \hline\hline
2.1 & 0.25& 1.35 & 1.20 &   2.00 & 1.61 \\ \hline
2.1 & 0.5 & 1.50 & 1.40 &   2.50 & 2.20 \\ \hline
2.1 & 1   & 1.69 & 1.65 &   3.50 & 3.27 \\ \hline
2.1 & 1.5 & 1.81 & 1.78 &   4.50 & 4.32 \\ \hline
2.1 & 2   & 1.88 & 1.87 &   5.50 & 5.35 \\ \hline
\end{tabular}

\bigskip
\noindent \textbf{Table 2: }Same as Table 1 for double strand DNA.
\medskip

\noindent 
\begin{tabular}{|c|c|c|c|c|c|}
\hline
$\labare \ell _{B}$ & ~~$\kappa a$~~ & 
~~$\laeff\ell _{B}\quad [Eq. (\ref{EffchargeCyl})]$~~ & 
~~$\laeff\ell _{B}\quad (num)$~~ & 
~~$\laeff^{\sat}\ell _{B}\quad [Eq. (\ref{EffchargeSatCyl})]$~~ & 
~~$\laeff^{\sat}\ell _{B}\quad (num)$~~ \\ \hline\hline
4.2 & 0.25& 1.65 & 1.41 &   2.00 & 1.61 \\ \hline
4.2 & 0.5 & 1.94 & 1.78 &   2.50 & 2.20 \\ \hline
4.2 & 1   & 2.40 & 2.31 &   3.50 & 3.27 \\ \hline
4.2 & 1.5 & 2.76 & 2.71 &   4.50 & 4.32 \\ \hline
4.2 & 2   & 3.02 & 2.99 &   5.50 & 5.35 \\ \hline
\end{tabular}


\begin{thebibliography}{99}
\bibitem{Kjellander} R. Kjellander in: Proceedings of the NATO Advanced
Study Institute on Electrostatic Effects in Soft Matter and Biophysics, ed.
by C. Holm et al., Kluwer, p317 (2001).

\bibitem{Belloni} L. Belloni, Colloids Surfaces A: Physicochem. Eng. Aspects 
\textbf{140}, 227 (1998).

\bibitem{Hansen} J.-P. Hansen and H. L\"{o}wen, Annu. Rev. Phys. Chem. 
\textbf{51}, 209 (2000).

\bibitem{Levin} Y. Levin, Rep. Prog. Phys. {\bf 65} 1577 (2002). 

\bibitem{Lukatsky} D.B. Lukatsky and S.A. Safran, 
Phys. Rev. E {\bf 63}, 011405 (2001). 

\bibitem{Alexander} S. Alexander, P.M. Chaikin, P. Grant, G.J. Morales and
P. Pincus, J. Chem. Phys. \textbf{80}, 5776 (1984).

\bibitem{Katchalsky} R.M. Fuoss, A. Katchalsky and S. Lifson, Proc. Natl.
Acad. Sci. U.S.A. \textbf{37}, 579 (1951). See also T. Alfrey, P. Berg and
H.J. Morawetz, J. Polym. Sci. \textbf{7}, 543 (1951).

\bibitem{Manning} G.S. Manning, J. Chem. Phys. \textbf{51}, 924 (1969); F.
Oosawa, Polyelectrolytes, Dekker, New York (1971).

\bibitem{Letter} E. Trizac, L. Bocquet and M. Aubouy, 
Phys. Rev. Lett.  {\bf 89}, 248301 (2002). 

\bibitem{JCP} L. Bocquet,  E. Trizac and M. Aubouy,
J. Chem. Phys. {\bf 117}, 8138 (2002).

\bibitem{Rque} For alternative definitions, see e.g. A. Diehl, M.C. Barbosa
and Y. Levin, Europhys. Lett. \textbf{53}, 86 (2001); M. Deserno, C. Holm
and S. May, Macromolecules \textbf{33}, 199 (2000).

\bibitem{Shkel} I.A. Shkel, O.V. Tsodirov and M.T. Record, J. Phys. Chem. B, 
\textbf{104}, 5161 (2000).

\bibitem{Groot} R.D. Groot, J. Chem. Phys. \textbf{95}, 9191 (1991).

\bibitem{Stevens} M.J. Stevens, M.L. Falk and M.O. Robbins, J. Chem. Phys. 
\textbf{104}, 5209 (1996).

\end{thebibliography}
\end{document}